\documentclass{osa-article}
\usepackage{xcolor}
\journal{osac}

\articletype{Research Article}

\begin{document}

\title{Negative cavity photon spectral function in an optomechanical system with two parametrically-driven mechanical modes}

\author{Ali Motazedifard,\authormark{1,2,$ \dagger $} A. Dalafi,\authormark{3,*} and M. H. Naderi\authormark{1,4,$ \dagger  \dagger$ }}

\address{\authormark{1}Quantum Optics Group, Department of Physics, University of Isfahan, Hezar-Jerib, 81746-73441, Isfahan, Iran\\
	\authormark{2}Quantum Sensing Lab, Quantum Metrology Group, Iranian Center for Quantum Technologies (ICQT), Tehran, Iran\\
	\authormark{3}Laser and Plasma Research Institute, Shahid Beheshti University, Tehran, Iran\\
	\authormark{4}Department of Physics, University of Isfahan, Hezar-Jerib, 81746-73441, Isfahan, Iran\\
	
	\authormark{$ \dagger $}motazedifard.ali@gmail.com\\
	\authormark{$ \dagger \dagger $} mhnaderi@phys.ui.ac.ir\\
	\authormark{*}a\_dalafi@sbu.ac.ir\\
}




\begin{abstract}
	We propose an experimentally feasible optomechanical scheme to realize a negative cavity photon spectral function (CPSF) which is equivalent to a negative absorption. The system under consideration is an optomechanical system consisting of two mechanical (phononic) modes which are linearly coupled to a common cavity mode via the radiation pressure while parametrically driven through the coherent time-modulation of their spring coefficients. Using the equations of motion for the cavity retarded Green's function obtained in the framework of the generalized linear response theory, we show that in the red-detuned and weak-coupling regimes a frequency-dependent effective cavity damping rate (ECDR) corresponding to a negative CPSF can be realized by controlling the cooperativities and modulation parameters while the system still remains in the stable regime. Nevertheless, such a negativity  which acts as an optomechanical gain never occurs in a standard (an unmodulated bare) cavity optomechanical system. Besides, we find that the presence of two modulated mechanical degrees of freedom provides more controllability over the magnitude and bandwidth of the negativity of CPSF, in comparison to the setup with a single modulated mechanical oscillator. Interestingly, the introduced negativity may open a new platform to realize an extraordinary (modified) optomechanically induced transparency (in which the input signal is amplified in the output) leading to a perfect tunable optomechanical filter with switchable bandwidth  which can be used as an optical transistor.
\end{abstract}

\section{Introduction}
In the recent two decades, the field of quantum optomechanics \cite{Aspelmeyer}, in which a mechanical oscillator (MO) is linearly or quadratically coupled to an optical/microwave mode via the radiation pressure, using state-of-the-art technologies has significantly progressed both in theory and experiment. The optomechanical systems (OMSs) play a key role in quantum technologies and also can be employed as a controllable setup for the emergence of quantum effects on macroscopic scales or probing the fundamentals of physics \cite{optomechanicalBelltest1,pikovski2012Planckprobing}.

OMSs have been applied in a wide range of applications including ultraprecision quantum sensing and measurements \cite{xsensing1,CQNCPRL,aliNJP,complexCQNC,CQNCNatureexp,masssensing,allahverdiHomodyne,conditionalbackactionevading,aliDCEforcesenning,sillanpaasensing1,sillanpaasensing2,aliAVSbook2020,mehryMagnetometry2021}, quantum illumination or quantum radar \cite{quantumillumination1,quantumillumination2}, cooling of the MO \cite{groundstatecooling,sidebandcooling,lasercooling}, generation of entanglement \cite{Palomaki2,Paternostro,genesentangelment,dalafiQOC,barzanjehentanglement,foroudcrystalentanglement}, synchronization of MOs \cite{Mari1,MianZhang,Bagheri,foroudsynch}, generation of quadrature squeezing and amplification \cite{Clerkdissipativeoptomechanics,pontinmodulation,bothner2020,optomechanicswithtwophonondriving,clerkfeedback,Harris,Bowen,twofoldsqueezing,aliDCEsqueezing,Sillanpaa1,Sillanpaa2,Sillanpaa3,Sillanpaa4,Sillanpaa5,Wollman}, realizing the dynamical Casimir effect-based nonclassical radiation sources \cite{aliDCE1,aliDCE2,aliDCE3, NoriDCE1,NoriDCE2,NoriDCE3}, optomechanically induced transparency (OMIT) \cite{OMIT1,OMIT2,OMIT3,vitaliOMIT,marquardtOMIT,XiongOMIT,shahidaniOMIT2013} which is analogous to the familiar phenomenon of electromagnetically induced transparency (EIT) \cite{EIT}, optomechanically induced gain (OMIG) \cite{mikaeili2023slowlight}, realizing ultraslow light \cite{dalafiOMIT2022}, and also the realization of microwave nonreciprocity, unidirectional transport, isolator and circulator \cite{Barzanjehnonreciprocity}.


Very recently, by applying the generalized linear response theory (GLRT) to a standard driven-dissipative OMS \cite{aliGreen2021} as an open quantum system, the Green's functions equations of motion have been obtained from quantum Langevin equations (QLEs) in the Heisenberg picture to deeply explain a wealth of phenomena, including the anti-resonance, normal mode splitting (NMS), and OMIT. 
On the other hand, driven-dissipative quantum systems have recently attracted intensive research interests, particularly in the context of quantum control (for a very recent, comprehensive review, the reader is referred to Ref.~\cite{quantumcontrol}). 
Generally, the steady states of such systems can be far from equilibrium because of the unavoidable competition between driving and dissipation. The steady-state behaviors of such systems have been investigated theoretically by many authors (see, for example, \cite{noh2017,petruccione2016,schiro2021}). In addition to the steady-state behavior, the description of the system's response to weak external perturbations is of great interest. It is well-known that the Green's function method is a powerful technique for this purpose. To determine the Green's functions of open quantum systems there are some approaches in the literature \cite{Arrigoni, Dorda, greenScarlatella1, greenScarlatella2,keldyshReview2016,banPRA17,banQStud15,shenPRA17} which are mainly based on the Lindblad master equation, i.e., presented in the Schr\"{o}dinger picture.

Motivated by the above-mentioned explanations, in this paper we are going to obtain the linear response of an OMS, consisting of two parametrically-driven mechanical modes coupled to a common electromagnetic field, to an input time-dependent probe field in the red-detuned and weak-coupling regimes. We show that under special conditions such a system exhibits an optomechanical gain which leads to the amplification of input probe with a controllable bandwidth window. For this purpose, we use the GLRT \cite{aliGreen2021} to calculate the output cavity field and show that it includes a carrier wave together with two sidebands whose amplitudes are obtained in terms of the cavity retarded Green's function (CRGF). In this way, the response of the system to the input probe field is obtained in terms of a special function which is called the CPSF. We show that by controlling the system parameters a negative CPSF may be achieved in the stable regime of the system which leads to the interesting phenomenon of extraordinary OMIT. In this situation the OMS behaves as an optical \textit{transistor} \cite{transistor1,transistor2,transistor3} exhibiting an optomechanical \textit{gain} which can amplify the input signal considerably. 
The advantage of an OMS with two parametrically driven MOs is that it provides more controllability to achieve more negative CPSF in the stable regime, and also enables us to manipulate the OMIT properties in comparison to an OMS with a single mechanical mode. Besides, It should be pointed out that unlike other standard systems such as the degenerate parametric amplifier (DPA), the occurrence of a negative CPSF in the present system takes place in the stable regime. We also show that the CPSF negativity is originated by a negative ECDR through the coherent time modulation of the mechanical modes.


The paper is organized as follows. In Sec.~\ref{sec2}, we describe the basic model of the system under study and its effective Hamiltonian.
In Sec.~\ref{sec3}, by solving the QLEs in frequency space we find the self-energies and induced-parametric squeezing coefficients corresponding to the induced-effective damping rates.
Then, in Sec.~\ref{glrt} we use the GLRT to obtain the linear response of the system to the input probe and define the CPSF.
Using the solutions in the frequency space, we calculate the CPSF in Sec.~\ref{negative SF}, and show that by controlling the cooperativities and modulation parameters it can be negative when the ECDR becomes negative. In Sec.~\ref{sec_experiment}, we present an experimental discussion which shows the possibility of the experimental realization of our theoretical proposal.
In Sec.~\ref{sec.comparisontotheOPA}, we present a physical interpretation of the CPSF negativity in our system together with a comparison to an ordinary optical parametric amplifier (OPA). 
Finally, we summarize our concluding remarks and outlooks in Sec.~\ref{summary}. In particular, the manifestation of the negative CPSF in the modified OMIT, negative effective temperature (NET), and slow-light as outlooks are briefly discussed in Sec.~\ref{summary}.

\section{System Hamiltonian}\label{sec2}

\begin{figure}
	\includegraphics[width=7cm]{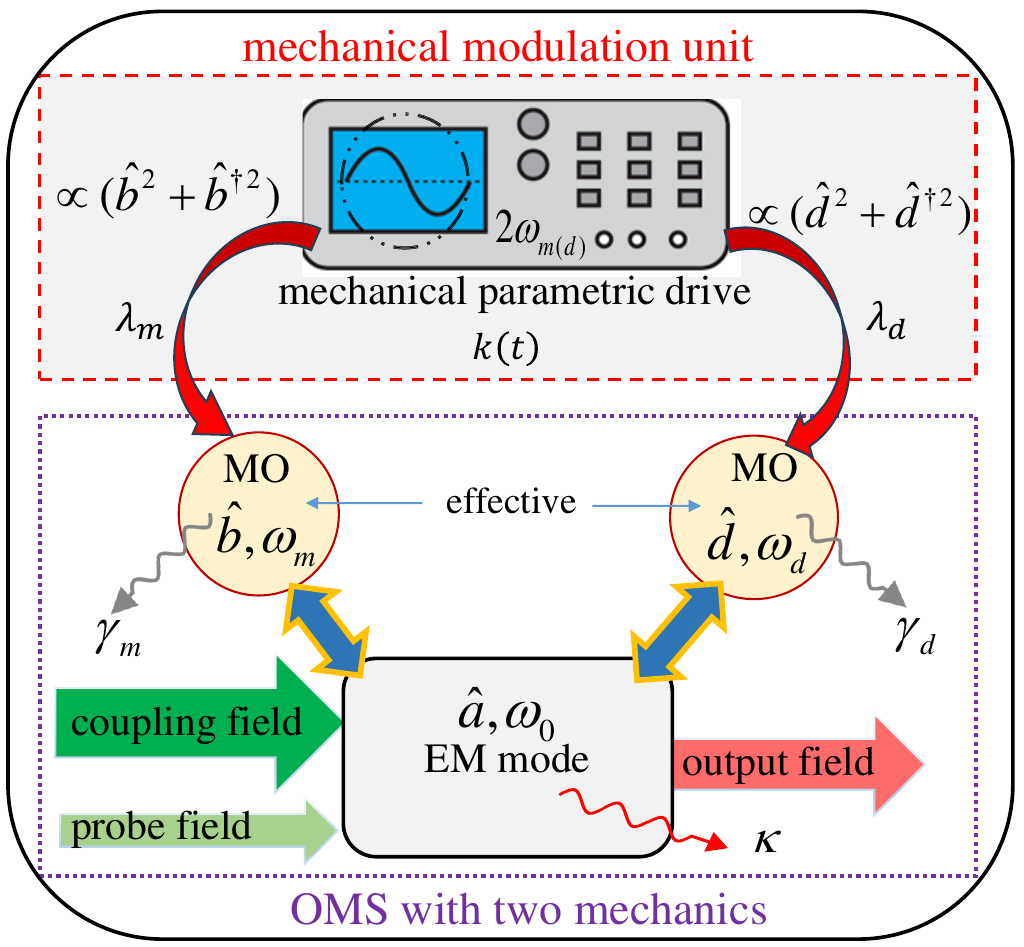}
	\centering
	\caption{Schematic of a generic optomechanical system consisting of two similar effective MOs which are linearly coupled to a common electromagnetic (EM) mode via the radiation pressure. The spring coefficients of the MOs are parametrically driven at twice their natural frequencies so that $ k(t)=k_{m(d)}+\delta k_{m(d)} \sin (2\omega_{m(d)}t +\varphi_{m(d)}) $. The cavity mode is driven by two coupling and probe classical coherent lasers such that the probe power is much weaker than both the coupling and modulating powers.
	}
	\label{fig1}
\end{figure}

As depicted in Fig.~(\ref{fig1}), we consider  a generic OMS consisting of two effective MOs with the natural frequencies $ \omega_{m} $ and $ \omega_{d} $, and damping rates $ \gamma_{m} $ and $ \gamma_{d} $ interacting with the radiation pressure of the single optical mode of the cavity while the spring coefficients of the MOs are coherently modulated at twice their natural frequencies. Besides, the natural frequency of the optical mode inside the cavity and the cavity damping rate are, respectively, $ \omega_0 $ and $ \kappa=\kappa_e +\kappa_i$, where $\kappa_i$  and $\kappa_e$ are, respectively, the internal and external decay rates. It is also assumed that the cavity is driven at the rate of $ \varepsilon_c=\sqrt{\kappa_{e} P_c / \hbar \omega_c}  $ by a strong coupling laser with frequency $\omega_{c}$ and the input power $ P_c $. The total Hamiltonian of the system in the frame rotating at the coupling laser frequency, $ \omega_c $, can be written as \cite{aliDCE3,optomechanicswithtwophonondriving}
\begin{eqnarray} \label{H1total}
	&&  \hat H= \hbar \Delta_c \hat a^\dag \hat a +\hbar \omega_m \hat b^\dag \hat b + \hbar \omega_d \hat d^\dag \hat d + i\hbar\varepsilon_c (\hat a^\dag - \hat a)  -\hbar g_0 \hat a^\dag \hat a (\hat b + \hat b^\dag) -\hbar G_0 \hat a^\dag \hat a (\hat d + \hat d^\dag) \nonumber \\
	&& \qquad \qquad +\frac{i \hbar}{2} (\lambda_m \hat b^{\dag 2}  e^{-2i\omega_m t}- \lambda_m^\ast \hat b^2 e^{2i\omega_m t}) +\frac{i \hbar}{2} (\lambda_d \hat d^{\dag 2}  e^{-2i\omega_d t}- \lambda_d^\ast \hat d^2 e^{2i\omega_d t}).
\end{eqnarray}

The first four terms in the Hamiltonian describe, respectively, the free energy of the optical mode, the free energy of the MOs and the coupling between the optical mode and the classical driving laser. Here, $ \hat a $, and $ \hat b $($ \hat d $) are the annihilation operators of the optical and mechanical modes, and $ \Delta_c= \omega_0-\omega_c $ is the detuning of the optical mode from the coupling laser frequency. The fifth and sixth terms denote, respectively, the optomechanical interactions between the optical mode and the two mechanical modes $ \hat b $ and $ \hat d $ with the single-photon optomechanical coupling strengths $ g_0$ and $ G_0 $. 

The seventh and eighth terms account for the parametric driving of the MOs spring coefficients at the twice of their natural frequencies giving rise to the time-dependent spring coefficient as $ k_{m(d)}(t)=k_{m(d)}+\delta k_{m(d)} \sin (2\omega_{m(d)} t + \varphi_{m(d)}) $ where $ \varphi_{m(d)} $ is the phase of external modulation (for more details see Appendix.~A). These terms have been written in the rotating wave approximation (RWA) which is valid over the time scales longer than $ \omega_{m(d)}^{-1} $. Here, $ \lambda_{m(d)}= \vert \lambda_{m(d)} \vert e^{-i\varphi_{m(d)}} $ is the mechanical modulation amplitude with $ \vert \lambda_{m(d)} \vert= \delta k_{m(d)} x_{zp(m(d))}^2 / 2\hbar $ \cite{aliDCE3,optomechanicswithtwophonondriving} where $ x_{zp(m(d))}= \sqrt{\hbar / 2m_{m(d)}\omega_{m(d)}} $ is the zero-point position fluctuation. Note that by fixing the phase of modulation, it is always possible to take $ \lambda_{m(d)} $ as a real number. 
It is worth mentioning that the parametric mechanical driving term can be considered as the mechanical phonon analog of the degenerate parametric amplification which may lead to the dynamical Casimir effect of mechanical phonons \cite{aliDCE3}. Furthermore, the parametric driving has been recently used for the preparation of an optomechanical system in a two-mode squeezed thermal state \cite{pontinmodulation}.It is pointed out that the parametric mechanical driving can be realized in
a superconducting microwave optomechanical circuit by applying a combination of a static voltage and an oscillating voltage (ac-voltage) in order to simulate the modulation of the spring coefficient \cite{bothner2020}.

We keep our discussion general because there are several different experimental realizations for the scheme proposed here. As some experimental setups that can be matched to our scheme one can consider an OMS consisting of: (1) two membranes-in-the-middle inside an optical cavity, (2) one moving-end mirror and a membrane in the middle, (3) two moving-end mirrors, (4) one membrane-in-the-middle with two internal mechanical (surface) modes, (5) a single-mode Bose-Einstein condensate inside an OMS with a moving-end mirror where the Bogoliubov mode of the BEC plays the role of the second mechanical mode, and (6) a two-mode BEC inside an optical cavity where the two collective modes of the BEC play the role of the two mechanical modes. However, it should be noted that throughout the paper we will use feasible ratios for dimensionless system parameters which can be obtained in any above mentioned setups as the controllable parameters.

In the continuation, we are going to find the linear response of the OMS described by the Hamiltonian of Eq.(\ref{H1total}) to a weak time-dependent perturbation arising from a weak probe laser with frequency $\omega_p$ which interacts with cavity through the following potential in the frame rotating at the coupling laser frequency
\begin{equation}\label{Vt}
	\hat V(t)=i\hbar\varepsilon_{p}^\ast \hat a e^{i\omega_{pc} t} - i\hbar\varepsilon_{p} \hat a^\dagger e^{-i\omega_{pc} t},
\end{equation}
where $\omega_{pc}=\omega_p-\omega_c$ being the detuning between the probe and the coupling laser frequencies. Here, it is assumed that the probe laser drives the cavity at a rate of $\varepsilon_p$ whose absolute value is much smaller than the coupling laser pump rate and the modulation amplitudes ($|\varepsilon_p|\ll\varepsilon_c, \lambda_m, \lambda_d$) so that the potential of Eq.(\ref{Vt}) can be considered as a weak time-dependent perturbation against the system Hamiltonian of Eq.(\ref{H1total}). In order to find the linear response of the OMS to the probe perturbation, we make use of the GLRT described in Ref\cite{aliGreen2021}. Based on the GLRT the linear response of a driven-dissipative quantum system is described by the open system Green's function in the absence of the perturbation. Therefore, we first study the dynamics of the system in the absence of perturbation in section \ref{sec3}, and then we obtain the linear response of the system to the perturbation using the non-perturbative Green's functions described by the GLRT in section \ref{glrt}.

\section{Dynamics of the system in the absence of the probe perturbation \label{sec3}}

Due to the nonlinearity and time-dependence of the Hamiltonian of Eq.(\ref{H1total}), the corresponding QLEs are nonlinear operator equations with time-dependent coefficients which are not solvable. In order to make them solvable, the first step is to linearize the nonlinear QLEs which is valid as far as the quantum fluctuations are much smaller than the mean-fields. However, the linearized QLEs of the present system are not still solvable due to the existence of time-dependent coefficients. Nevertheless, if the two MOs have equal natural frequencies ($ \omega_m=\omega_d $), in the interaction picture, and in the so-called red detuned regime under the RWA, the system dynamics has an analytical solution. Interestingly, many important phenomena become observable in the red detuned regime of optomechanics because the system is on-resonance.

Therefore,  based on the above explanations, in the red-detuned regime of $ \Delta_0 = \omega_m = \omega_d $, in which $ \Delta_{0}=\Delta_c-2g_0 \bar b-2G_0 \bar d $  is the effective cavity detuning and $ \bar b\approx g_0 \bar a^2/\omega_{m} $ and $ \bar d\approx -G_0 \bar a^2/\omega_{d} $ are the mean mechanical fields with $\bar a=\varepsilon_c/\sqrt{\kappa^2/4+\Delta_{0}^{2}} $ being the steady-state mean value of the optical mode, the linearized QLEs for the quantum fluctuations of the system operators can be obtained in the interaction picture and under RWA as the following set of equations \cite{aliDCE3}
\begin{subequations}
	\begin{eqnarray}
		&& \!\!\!\!\! \delta \dot { \hat a} = -\frac{\kappa}{2} \delta \hat a + i g  \delta \hat b + i G \delta \hat d + \sqrt{\kappa} \hat a_{in} , \label{QLEBS a} \\
		&&\!\!\!\!\! \delta \dot {\hat b}  = -\frac{\gamma_m}{2} \delta \hat b + i g \delta \hat a + \lambda_m \delta \hat b^\dag +  \sqrt{\gamma_m}\hat b_{in}, \label{QLEBS b} \\
		&&\!\!\!\!\! \delta \dot {\hat d}  = - \dfrac{\gamma_d}{2} \delta \hat d  +i G  \delta \hat a + \lambda_d \delta \hat d^\dag  +  \sqrt{\gamma_d} \hat d_{in}, \label{QLEBS c}
	\end{eqnarray}
\end{subequations}
where $ g=g_0\bar a $ and $G=G_0\bar a$ are the enhanced-optomechanical coupling strengths. Moreover, the optical and the Brownian mechanical input noises $ \hat o_{in} $ ($ o=a,b,d $) satisfy the Markovian correlation functions  $ \langle \hat o_{in}(t) \hat o_{in}^\dag (t') \rangle= (1+\bar n_j^T) \delta (t-t') $ and $ \langle \hat o_{in}^\dag (t) \hat o_{in}(t') \rangle= \bar n_j^T \delta (t-t') $ ($ j=c,m,d $) where $ \bar n_j^T=[{\rm exp}(\hbar \omega_j /k_B T)-1]^{-1} $ is the mean number of excitations corresponding to thermal optical and mechanical reservoir at temperature $ T $.

The linearized QLEs of motion, i.e., Eqs.(\ref{QLEBS a}-\ref{QLEBS c}) together with their Hermitian conjugates can be rewritten in the following compact form
\begin{equation} \label{u1}
	\frac{d}{dt} \hat{ \boldsymbol{u}}(t)= \boldsymbol{\chi}_0 \hat{ \boldsymbol{u}}(t) + \hat{ \boldsymbol{u}}_{in}(t), 
\end{equation}
where $ \hat{ \boldsymbol{u}}(t)=(\delta\hat a,\delta \hat a^\dagger, \delta\hat b, \delta\hat b^\dagger, \delta\hat d, \delta\hat d^\dag)^{\rm T} $ is the vector of the quantum field fluctuations and $  \hat{ \boldsymbol{u}}_{in}(t)= (\sqrt{\kappa} \hat a_{in},\sqrt{\kappa} \hat a^{\dag}_{in},\sqrt{\gamma_m} \hat b_{in}, \sqrt{\gamma_m} \hat b^{\dag}_{in}, \sqrt{\gamma_d} \hat d_{in}, \sqrt{\gamma_d} \hat d^{\dag}_{in} )^{\rm T} $ is the vector of quantum noises. The drift matrix $ \boldsymbol{\chi}_0  $ can be found easily from the equations of motion as follows
\begin{eqnarray} \label{chi0parametricOMS}
	&&\!\!\!\!\!\!\!\!\!\!\!\!  \boldsymbol{\chi}_0 \!=\! \left( \begin{matrix}
		{-\frac{\kappa}{2}} & {0} & {ig} & {0} & {iG} & {0}  \\
		{0} & {-\frac{\kappa}{2}} & {0} & {-ig} & {0} & {-iG}  \\
		{ig} & {0} & {-\frac{\gamma_m}{2}} & {\lambda_m} & {0} & {0}  \\
		{0} & {-ig}& {\lambda_m^\ast} & {-\frac{\gamma_m}{2}} & {0} & {0}   \\
		{iG} & {0} & {0} & {0} & {-\frac{\gamma_d}{2}} & {\lambda_d}  \\
		{0} & {-iG} & {0} & {0} & {\lambda_d^\ast} & {-\frac{\gamma_d}{2}}  \\
	\end{matrix} \right).
\end{eqnarray}

It is obvious that one can solve Eq.(\ref{u1}) in the Fourier space as follows
\begin{eqnarray} \label{u_w}
	&& \hat{ \boldsymbol{u}}(\omega)= \boldsymbol{\chi}(\omega) \hat{ \boldsymbol{u}}_{in}(\omega), 
\end{eqnarray}
where the susceptibility matrix $  \boldsymbol{\chi}(\omega)  $ is generally defined as
\begin{eqnarray}\label{chiw}
	&&  \boldsymbol{\chi}(\omega) = \Big(-i\omega\boldsymbol{I} -  \boldsymbol{\chi}_0 \Big)^{-1},
\end{eqnarray}
in which $\boldsymbol{I}$ is the $6\times 6$ identity matrix. Now, based on Eq.(\ref{u_w}) the Fourier transform of the optical field can be obtained as follows 
\begin{equation} \label{a_w}
	\delta \hat a(\omega)= \sum_{o=a,b,d} \sqrt{\kappa_o} \Big[\chi_{ao}(\omega) \hat o_{in}(\omega) + \chi_{ao^\dag}(\omega) \hat o^\dag_{in}(\omega)\Big], \\
\end{equation}
with $\chi_{ij}(\omega)$ being the susceptibility matrix elements, the expressions of which will be given later [Eqs.(\ref{chiparametric a}-\ref{chiparametric f})]. Note that in the above equation $ \kappa_a \equiv \kappa $, $ \kappa_b \equiv \gamma_m $, and $ \kappa_d \equiv \gamma_d $.


In the following, in order to compare the present modulated optomechanical system with the well-known DPA we write the solution to the Eq.~(\ref{u1}) for the the optical field noise operator and its Hermitian conjugate as
\begin{subequations} \label{a_omega_eq}
	\begin{eqnarray} 
		-i\omega \delta \hat a(\omega)&=&-\Big(\kappa / 2 + i \Sigma_a(\omega)\Big) \delta \hat a(\omega) + \tilde \lambda_a(\omega) \delta \hat a^\dag(\omega) + \sqrt{\kappa} \hat A_{in}(\omega), \label{a_dasg a}\\
		-i\omega \delta \hat a^\dag(\omega) \! &=&-\Big(\kappa / 2 - i \Sigma_a^\ast(-\omega)\Big) \delta \hat a^\dag(\omega) + \tilde \lambda_a^\ast(-\omega) \delta \hat a(\omega) + \sqrt{\kappa} \hat A_{in}^\dag(\omega),\label{a_dasg b}
	\end{eqnarray}
\end{subequations}
where the cavity self-energy $ \Sigma_a(\omega) $ and the induced frequency-dependent parametric amplification coefficient $\tilde\lambda_a(\omega) $, which is analogous to the gain factor in the conventional OPA, are, respectively, given by \cite{aliDCE3}
\begin{subequations} \label{lambda_all}
	\begin{eqnarray}
		&&\!\!\!\!\!\! i \Sigma_a(\omega)=\! g^2 \frac{\frac{\gamma_m}{2}-i \omega}{(\frac{\gamma_m}{2}-i \omega)^2-\vert \lambda_m \vert^2} \! + \! G^2 \frac{\frac{\gamma_d}{2}-i \omega}{(\frac{\gamma_d}{2}-i \omega)^2-\vert \lambda_d \vert^2} , \\
		&&\!\!\!\!\!\!\!\!\! \tilde \lambda_a(\omega)=\frac{ g^2  \lambda_m}{(\gamma_m/2-i \omega)^2-\vert \lambda_m \vert^2} + \frac{G^2 \lambda_d}{(\gamma_d/2-i \omega)^2-\vert \lambda_d \vert^2}. \label{lambda_a_omega}
	\end{eqnarray}
\end{subequations}
Furthermore, the generalized cavity input-noise in the frequency space is given by (see Ref.~\cite{aliDCE3})
\begin{align} \label{cavitynoiseA}
	\hat A_{in}=& \hat a_{in} + ig \sqrt{\frac{\gamma_m}{\kappa}} \frac{\gamma_m/2-i \omega}{(\gamma_m/2-i \omega)^2-\vert \lambda_m \vert^2} \left[ \hat b_{in} + \frac{\lambda_m}{\gamma_m/2-i\omega} \hat b^\dagger_{in}  \right] \nonumber \\
	& \qquad  +iG \sqrt{\frac{\gamma_d}{\kappa}} \frac{\gamma_d/2-i \omega}{(\gamma_d/2-i \omega)^2-\vert \lambda_d \vert^2} \left[ \hat d_{in} + \frac{\lambda_s}{\gamma_d/2-i\omega} \hat d^\dagger_{in}  \right] .
\end{align}
It should be noted that the susceptibility matrix of the system, $ \boldsymbol{\chi} $, can be found by rewriting the self-energies as \cite{aliDCEsqueezing} $ \boldsymbol{\Sigma}[\omega]= i (\boldsymbol{\chi}_0^{-1}[\omega] - \boldsymbol{\chi}^{-1}[\omega]) $ with $ \boldsymbol{\chi}_0^{-1}[\omega]=(\kappa/2-i \omega) \boldsymbol{I} $. As is seen, the cavity self-energy satisfies the relation $ \Sigma_a^\ast(-\omega)=- \Sigma_a(\omega)$. 
As is evident, by turning on the modulations, i.e., with $\lambda_{m,d} \neq 0$, the two mechanical modes do indeed mediate a parametric amplifier-like effective squeezing interaction for the optical mode which is also frequency-dependent (see Eq. (\ref{lambda_a_omega})), unlike the case of the conventional DPA.
Recently, it has been shown \cite{aliDCEsqueezing} that the mechanical parametric drive can lead to a strong and robust optomechanical squeezing and phase-sensitive amplification which can provide a situation for an ultraprecision single-quadrature force sensing \cite{aliDCEforcesenning}. 
Moreover, as is clear from Eq.~(\ref{lambda_a_omega}), for $ \lambda_{m(d)}^\ast=\lambda_{m(d)} $ (real parametric amplifications (paramps)) one has $ \tilde \lambda_a^\ast(-\omega)=\tilde \lambda_a(\omega) $.

By solving the coupled Eqs.~(\ref{a_dasg a}) and (\ref{a_dasg b}), one can find the intracavity fluctuation operator $ \delta \hat a(\omega) $ in the following form
\begin{eqnarray} \label{asolve}
	&& \delta \hat a(\omega)= \sqrt{\kappa} \chi_{aa}(\omega) \left[ \hat A_{in}(\omega)+ \mathcal{M}(\omega) \hat A_{in}^\dag(\omega) \right] ,
\end{eqnarray}
where
\begin{eqnarray} \label{momega}
	&& \mathcal{M}(\omega)= \frac{\tilde \lambda_a(\omega)}{\kappa/2-i\Big(\omega +\Sigma_a^\ast(-\omega)\Big)},
\end{eqnarray}
and the matrix element $\chi_{aa}(\omega)$ is given by Eq.(\ref{chiparametric a}) in which the modified cavity self-energy is as follows
\begin{eqnarray} \label{sigmatilde}
	&& \tilde \Sigma_a(\omega)=\Sigma_a(\omega)-\frac{\tilde \lambda_a(\omega) \tilde \lambda_a^\ast(-\omega) }{i\kappa/2+\Big(\omega+\Sigma_a^\ast (-\omega)\Big)} \quad. \label{finalselfenergy}
\end{eqnarray}
It is straightforward to show that for $ \lambda_{m(d)}^\ast=\lambda_{m(d)} $, we have $ \tilde \Sigma_a^\ast(-\omega)= - \tilde \Sigma_a(\omega)$ and $ \mathcal{M}^\ast(-\omega)=\mathcal{M}(\omega) $. 
The elements of the susceptibility matrix $\boldsymbol{\chi}(\omega)$ which are derived from the solution to Eq.~(\ref{u1}) in the frequency space, read as
\begin{subequations}
	\begin{eqnarray}
		&& \chi_{aa}(\omega)= \frac{i}{i\kappa/2 + \Big(\omega - \tilde \Sigma_a(\omega)\Big)} \quad,\label{chiparametric a}\\
		&& \chi_{aa^\dag}(\omega)= \chi_{aa}(\omega) \mathcal{M}(\omega), \label{chiparametric b}\\
		&&  \chi_{ab}(\omega)=\chi_{aa}(\omega) \frac{ig \left[(\frac{\gamma_m}{2} - i \omega) - \lambda_m^\ast \mathcal{M}(\omega)\right]}{(\frac{\gamma_m}{2} - i \omega)^2 - \vert \lambda_m \vert^2} , \label{chiparametric c}\\
		&& \chi_{ab^\dag}(\omega)=\chi_{aa}(\omega) \frac{ig \left[ \lambda_m- (\frac{\gamma_m}{2} - i \omega) \mathcal{M}(\omega)\right]}{(\frac{\gamma_m}{2} - i \omega)^2 - \vert \lambda_m \vert^2} , \label{chiparametric d}\\
		&& \chi_{ad}(\omega)=\chi_{aa}(\omega) \frac{iG \left[(\frac{\gamma_d}{2} - i \omega) - \lambda_d^\ast \mathcal{M}(\omega)\right]}{(\frac{\gamma_d}{2} - i \omega)^2 - \vert \lambda_d \vert^2} , \label{chiparametric e}\\
		&& \chi_{ad^\dag}(\omega)=\chi_{aa}(\omega) \frac{iG \left[ \lambda_d- (\frac{\gamma_d}{2} - i \omega) \mathcal{M}(\omega)\right]}{(\frac{\gamma_d}{2} - i \omega)^2 - \vert \lambda_d \vert^2} .\label{chiparametric f}
	\end{eqnarray}
\end{subequations}
It can be easily shown that for real paramps, $ \lambda_{m(d)}=\lambda^\ast_{m(d)} $, one has $ \chi_{aa}^\ast(-\omega)=\chi_{aa}(\omega) $.

\section{Linear response of the OMS to the probe perturbation \label{glrt}}

In this section we investigate the linear response of the linearized modulated OMS described in Sec.~\ref{sec3} to the weak time-dependent probe perturbation of Eq.(\ref{Vt}) whose linearized form in the interaction picture and in the frame rotating at the coupling laser frequency is given by
\begin{equation}\label{dVt}
	\hat V(t)=i\hbar\varepsilon_{p}^\ast\delta \hat a e^{i\omega t}-i\hbar\varepsilon_{p}\delta \hat a^\dagger e^{-i\omega t},
\end{equation}
where $\omega=\omega_{pc}-\omega_m$.

Based on the GLRT \cite{aliGreen2021} the response of the intracavity optical field fluctuation to the time-dependent perturbation of Eq.(\ref{Vt}), in the interaction picture and in the frame rotating at the coupling laser frequency, is described by the following equation 
\begin{equation}
	\langle\delta \hat a(t)\rangle=\langle\delta \hat a\rangle_0+i\varepsilon_{p}^\ast\int_{-\infty}^{+\infty} dt^{\prime} G_{aa}^R(t-t^{\prime}) e^{i\omega t^{\prime}}
	-i\varepsilon_{p}\int_{-\infty}^{+\infty} dt^{\prime} G_{aa^\dag}^R(t-t^{\prime}) e^{-i\omega t^{\prime}},\label{mat}
\end{equation}
where $\langle\delta \hat a\rangle_0=0$ is the steady-state mean value of the optical field fluctuation in the absence of the time-dependent perturbation, and the open system retarded Green's functions are defined as \cite{aliGreen2021}
\begin{subequations}
	\begin{eqnarray}
		G_{aa}^R(t)=-i\theta(t)\langle [\delta\hat a(t),\delta\hat a(0)]\rangle_0,\label{Gaa}\\
		G_{aa^\dag}^R(t)=-i\theta(t)\langle [\delta\hat a(t),\delta\hat a^\dag(0)]\rangle_0\label{Gaad},
	\end{eqnarray}
\end{subequations}
where $\theta(t)$ it the Heaviside step function. The Green's functions of Eqs.(\ref{Gaa}-\ref{Gaad}) are called the cavity retarded Green's functions (CRGFs). It should be noted that the time evolutions of the operators in Eqs.(\ref{Gaa}-\ref{Gaad}) are obtained from the QLEs given by Eq.(\ref{u1}) derived in Sec.\ref{sec3}, and the subscript 0 means that all the expectation values should be calculated in the steady state of the system in the absence of the perturbation. On the other hand, Eq.(\ref{mat}) can be rewritten as 
\begin{equation}\label{mat-rfint}
	\langle\delta \hat a(t)\rangle=-i\varepsilon_{p} G_{aa^\dag}^R(\omega) e^{-i\omega t}+i\varepsilon_{p}^\ast G_{aa}^R(-\omega) e^{i\omega t},\\
\end{equation}
in terms of the Fourier transform of the Green's function, which is defined by $G(\omega)=\int_{-\infty}^{+\infty}d\tau G(\tau) e^{i\omega\tau}$.

Since $\langle \hat a(t)\rangle=\bar{a}+\langle\delta \hat a(t)\rangle$ is the expectation value of the optical field in the interaction picture and in the rotating frame, the response of the optical field to the time-dependent perturbation in the Heisenberg picture in the laboratory frame  is obtained as
\begin{equation}
	\langle \hat a(t)\rangle = \bar a e^{-i\omega_{c}t}-i\varepsilon_{p} G_{aa^\dag}^R(\omega_{pc}) e^{-i(\omega_{c}+\omega_{pc})t}+i\varepsilon_{p}^\ast G_{aa}^R(-\omega_{pc}) e^{-i(\omega_{c}-\omega_{pc})t},\label{Ra}\\
\end{equation}
As is seen from Eq.(\ref{Ra}), the optical mode has a central band (first term), the so-called carrier wave, oscillating with frequency $\omega_{c}$ and two sidebands, the so-called anti-Stokes (second term) oscillating with frequency $\omega_p$, and Stokes (third term) oscillating with frequency $2\omega_{c}-\omega_{p}$. As is well-known in optomechanics, in the resolved sideband regime ($\omega_m \gg \kappa$) if the coupling laser frequency is fixed at the red-detuned sideband of the cavity resonance ($\omega_c\approx\omega_0-\omega_m$) and the probe frequency scans around the neighborhood of cavity resonance ($\omega_p\approx\omega_0$), i.e., when $\omega_{pc}\approx\omega_m$, the anti-Stokes sideband (second term) is considerably enhanced  while the Stokes sideband (third term) is so much attenuated that becomes negligible [for more detail see Ref.\cite{aliGreen2021}].

Now, in order to obtain the response of the system to the input probe signal, one should use the input-output theory \cite{Aspelmeyer} which is given by $\varepsilon_{out}(t)+\varepsilon_{in}(t)=\kappa^\prime\langle \hat a(t)\rangle$, where $\kappa^{\prime}\ll\kappa$ is the rate at which the cavity is weakly coupled to the probe laser \cite{askari2021dynamics}. For simplicity, we go back to the interaction picture and rotating frame, in which the input probe signal is given by $\varepsilon_{in}(t)=-\varepsilon_p e^{-i\omega t}$, and use Eq.(\ref{mat-rfint}) to obtain the output (reflected) cavity field as
\begin{eqnarray}\label{eout}
	\varepsilon_{out}(t)=2\kappa^\prime\bar a + \Big(1-i\kappa^\prime  G_{aa^\dag}^R(\omega)\Big)\varepsilon_p e^{-i\omega t} + i\kappa^\prime\varepsilon_p^\ast  G_{aa}^R(-\omega) e^{i\omega t}.
\end{eqnarray}
Again, there is a central band and two sidebands in the output (reflected) field. Therefore, the reflected field amplitude at the probe frequency, which is defined as the ratio of the output response at the probe frequency (the second term in Eq.(\ref{eout})) to the input probe amplitude $\varepsilon_p$, is obtained as
\begin{equation}
	\varepsilon_r(\omega)=1-i\kappa^\prime G_{aa^\dag}^R(\omega),
\end{equation}
where the power reflection coefficient, defined as
\begin{eqnarray}
	\mathcal{R}(\omega)=|\varepsilon_r(\omega)|^2= 1-2\kappa^\prime {\rm Im}G_{aa^\dag}^R(\omega) + \kappa^{\prime 2} \vert G_{aa^\dag}^R(\omega)  \vert^2 ,
\end{eqnarray}
is the normalized power spectrum \cite{milburn2015quantumBook} of the output field at the frequency of the probe laser. It is approximately given by the following equation in the limit of $\kappa^\prime\ll\kappa$
\begin{equation}\label{R}
	\mathcal{R}(\omega)\approx 1-\kappa^\prime\mathcal{A}(\omega),
\end{equation}
where $\mathcal{A}(\omega)$, the so-called cavity photon spectral function, is defined as
\begin{equation}
	\mathcal{A}(\omega):=-2{\rm Im} G^R_{aa^\dag}(\omega) \label{A1},
\end{equation}
that is usually interpreted as an effective density of single-particle states\cite{optomechanicswithtwophonondriving,aliGreen2021}. Equations (\ref{R}) and (\ref{A1}) show simply that if $\mathcal{A}(\omega)>0$, then $\mathcal{R}(\omega)<1$ which means that the probe photon number in the output of the cavity is smaller than the input photons because a number of input photons are absorbed by the cavity due to finite value of single-photon states inside the cavity. On the other hand, in the case of \textit{ordinary} OMIT where $\mathcal{A}(\omega)=0$ at the center of transparency window, the cavity reflection coefficient is unity, i.e., all the input probe photons are completely reflected by the cavity because there is no single-photon state inside the cavity, and therefore the cavity cannot accommodate them inside. Finally, there is a very interesting case in which $\mathcal{A}(\omega)<0$, i.e., $\mathcal{R}>0$, which is the main subject of the present work. In this case, another kind of OMIT (the so-called \textit{extraordinary} OMIT) happens, where the output photons at the probe frequency becomes greater than input photons. From the physical point of view, the negativity of the cavity spectral function implies that the cavity acts as a photon generator which increases the number of the input photons at the output. In other words, under this \textit{optomechanical gain} condition the OMS behaves as an \textit{optical transistor} which can amplify the input (probe) signal. Therefore, it can be concluded that the spectral function is a measure of the absorptive response of the system against the external perturbation so that for $\mathcal{A}\ge0$ there is a finite or zero absorption while for $\mathcal{A}<0$, which can be interpreted as a negative absorption, a gain is manifested in the system. This interpretation is completely clarified in the next section where the spectral function is expressed in therms of the so-called cavity effective damping rate (see Eq.(\ref{A2})).

As is seen, the GLRT provides for us a suitable mathematical framework to describe the response of a driven-dissipative quantum system to a weak time-dependent perturbation using open system Green's functions. In this way, we can predict theoretically the response of the present modulated OMS to a weak input probe signal in the output cavity field through the Green's function $G^R_{aa^\dag}(\omega)$. Here, it is worth emphasizing that the spectral function has a two-fold figure of merit. From the physical point of view, it is indicative of an important physical quantity, the so-called density of single particle states, which can describe many interesting physical phenomena. From the mathematical point of view, it is related to the imaginary part of the Green's function of the dynamical system which is related to the absorption properties of the system and can be calculated rigorously from the dynamics of the system.

In the next section, we obtain CRGF through the equations of motion of open system Green's functions predicted by the GLRT, and investigate the conditions of the spectral function negativity while the system is in the stable regime.

\section{Cavity retarded Green's function: negative cavity photon spectral function}\label{negative SF}
Here, it should be noted that the time evolution of the quantum field fluctuations in the Green's function definitions of Eqs.(\ref{Gaa}-\ref{Gaad}) are obtained from the QLEs given by Eqs.~(\ref{QLEBS a}-\ref{QLEBS c}), and the expectation values have been defined in the steady state of the system.

In the following, we will calculate the CRGFs using their equations of motion which are obtained through the GLRT. To drive the equation of motion for $ G_{aa^\dag}^R(t) $ ($ G_{aa}^R(t) $), one should multiply each of the equations in Eq.~(\ref{u1}) by $\hat a^{\dagger}(0)$ $\big(\hat a(0)\big)$ on the left and on the right, subtract them from each other and then taking their mean values. In this way, the Green's functions equations of motion can be obtained as the following compact forms
\begin{subequations}
	\begin{eqnarray}
		&& \frac{d}{dt} \boldsymbol{G}^R_{a^\dag}(t)= -i\delta(t)\boldsymbol{V}_{a^\dag}+\boldsymbol{\chi}_0 \boldsymbol{G}^R_{a^\dag}(t),\label{GAd}\\
		&& \frac{d}{dt} \boldsymbol{G}^R_{a}(t)= +i\delta(t)\boldsymbol{V}_{a}+\boldsymbol{\chi}_0 \boldsymbol{G}^R_{a}(t),\label{GA}
	\end{eqnarray}
\end{subequations}
in which the six-dimensional Green's functions vectors are defined as
\begin{eqnarray}
	&& \boldsymbol{G}^R_{a^\dag}(t)= \Big(G^R_{aa^\dag}(t), G^R_{a^\dag a^\dag}(t), G^R_{ba^\dag}(t), G^R_{b^\dag a^\dag}(t), G^R_{da^\dag}(t), G^R_{d^\dag a^\dag}(t)\Big)^{\rm T}, \nonumber \\
	&& \boldsymbol{G}^R_{a}(t)= \Big(G^R_{aa}(t), G^R_{a^\dag a}(t), G^R_{ba}(t), G^R_{b^\dag a}(t),G^R_{da}(t), G^R_{d^\dag a}(t) \Big)^{\rm T}, \nonumber
\end{eqnarray}
and $ \boldsymbol{V}_{a^\dag}:= (1,0,0,0,0,0)^{\rm T} $ and $ \boldsymbol{V}_{a}:= (0,1,0,0,0,0)^{\rm T} $ are fixed six-dimensional vectors.

Now, by taking the Fourier transforms of Eqs.(\ref{GAd}) and (\ref{GA}) one can find the Green's function vectors in the Fourier space as
\begin{subequations}
	\begin{eqnarray}
		&&  \boldsymbol{G}^R_{a^\dag}(\omega)= -i \boldsymbol{\chi}(\omega) \boldsymbol{V}_{a^\dag},\label{GAdw}\\
		&&  \boldsymbol{G}^R_{a}(\omega)= +i \boldsymbol{\chi}(\omega) \boldsymbol{V}_{a},\label{GAw},
	\end{eqnarray}
\end{subequations}
where $\boldsymbol{\chi(\omega)}$ is the susceptibility matrix defined by Eq.(\ref{chiw}). As is seen from Eqs.(\ref{GAdw}) and (\ref{GAw}), the CRGFs in the frequency space, i.e., the Fourier transform of Eqs.(\ref{Gaa}) and (\ref{Gaad}) can be obtained as
\begin{eqnarray}
	&& G^R_{aa^\dag}(\omega)=-i \chi_{aa}(\omega)=\frac{1}{i\kappa/2 + \Big(\omega - \tilde \Sigma_a(\omega)\Big)},\label{Gform} \\
	&& G^R_{ aa}(\omega)=-i \chi_{aa^\dag}(\omega)= G^R_{aa^\dag}(\omega) \mathcal{M}(\omega).
\end{eqnarray}
Note that for $ \lambda_{m(d)}^\ast=\lambda_{m(d)} $, it is easy to show that $  G^{R\ast}_{aa^\dag}(-\omega)=-  G^{R}_{aa^\dag}(\omega) $ and $  G^{R\ast}_{aa}(-\omega)=- G^{R}_{aa}(\omega) $.

It should be emphasized that the system Green's functions as well as the spectral function are independent of the input perturbation (input probe signal) while they depend on the system parameters in the absence of the perturbation (probe). It is one of the most important advantages of the linear response theory that can describe the dynamics of a nonequilibrium system (a system driven by a time-dependent perturbation) based on its nonperturbative properties (undriven system Green's functions). Furthermore, since the modified cavity self-energy is a complex valued function which can be written as $\tilde \Sigma_a(\omega)=\rm Re\tilde \Sigma_a(\omega)+i\rm Im \tilde \Sigma_a(\omega)$, the Green's function of Eq.(\ref{Gform}) can be written as
\begin{equation}
	G^R_{aa^\dag}(\omega)=\frac{1}{\Omega_{\rm eff}(\omega)+i\kappa_{\rm eff}(\omega)/2},
\end{equation} 
where $\Omega_{\rm eff}(\omega)=\omega-\rm Re \tilde \Sigma_a(\omega)$ and $\kappa_{\rm eff}(\omega)=\kappa- \kappa_{\rm opt}(\omega)$ with $ \kappa_{\rm opt}(\omega)=-2\rm Im \tilde \Sigma_a(\omega)$ are, respectively, the effective cavity frequency and effective cavity damping rate (ECDR). On the other hand, based on the definition of the CPSF of Eq.(\ref{A1}) it can be written in terms of the ECDR as the following form
\begin{equation}
	\mathcal{A}(\omega)=\frac{1}{|G^R_{aa^\dag}(\omega)|^2}\kappa_{\rm eff}(\omega)  \label{A2}.
\end{equation}
Based on Eq.(\ref{A2}), the negativity of the CPSF corresponds to the negativity of the ECDR ($\kappa_{\rm eff}(\omega)$) which is equivalent to the manifestation of a gain in the system while the positivity of the CPSF corresponds to a positive value of ECDR leading to the occurrence of dissipation. In short, it can be concluded that the spectral function is a measure of the absorptive response of the system to the input time-dependent perturbation whose negativity (positivity) corresponds to gain (dissipation).

In the following, we present all the results in the interaction picture and in the frame rotating at the coupling laser frequency in which $\omega=\omega_{pc}-\omega_m$, for simplicity. As is evident, the resonance condition of $\omega=0$ corresponds to $\omega_{pc}=\omega_m=\omega_d$.
We will show that in the present parametrically driven OMS the CPSF, $ \mathcal{A}(\omega) $, defined in Eq.(\ref{A1}) can be made negative by controlling the modulation parameters and cooperativities due to the parametric modulations of the phononic mechanical modes. The on-resonance CPSF can be easily calculated as follows
\begin{eqnarray}
	&& \mathcal{A}(0)=\frac{4}{\kappa}\frac{(\xi_m^2-1)(\xi_d^2-1)-\mathcal{C}_0(\xi_d^2-1)-\mathcal{C}_1(\xi_m^2-1)}{(\xi_m^2-1)(\xi_d^2-1)-2\mathcal{C}_0\mathcal{C}_1(\xi_m\xi_d-1)-\mathcal{C}_0(\mathcal{C}_0+2)(\xi_d^2-1)-\mathcal{C}_1(\mathcal{C}_1+2) (\xi_m^2-1)} \nonumber \\
	&& \qquad \qquad := \frac{4}{\kappa} \frac{A_N}{A_D}:=\frac{4}{\kappa} M'=\frac{1}{\kappa} M, \label{CPSF1}
\end{eqnarray}
where $\mathcal{C}_0=4g^2/\kappa \gamma_m$ and $\mathcal{C}_1= 4G^2/\kappa \gamma_d $ are the optomechanical cooperativities associated with the two mechanical modes $ \hat b $ and $ \hat d $, and $ \xi_{d(m)}=2\lambda_{d(m)}/\gamma_{d(m)}$ plays the role of an effective dimensionless amplitude of modulation. 
As is clearly seen, the on-resonance cavity photon spectral function ($\mathcal{A}(0)$) can be controlled by the effective amplitude of modulations $ \xi_{m,d} $ as well as the optomechanical cooperativities $ \mathcal{C}_{0} $ and $ \mathcal{C}_{1} $. It can be easily shown that the denominator $ A_D $ in Eq.~(\ref{CPSF1}) is always positive while the numerator $ A_N $ can be made negative in the stable regime.
As has been shown in Ref.~\cite{aliDCE3}, based on the Routh-Hurwitz criterion for the optomechanical stability condition, the modulation parameters $ \lambda_{m} $ and $ \lambda_{d} $ should satisfy the condition
\begin{eqnarray} \label{stability}
	&&\!\!\!\!\!\!\! \lambda_{m(d)} \le \gamma_{m(d)} \! + \! \Gamma_{op}^{m(d)}= \! \frac{\gamma_{m(d)}}{2} \left[1+ \mathcal{C}_{m(d)}\right],
\end{eqnarray}
where $ \Gamma_{op}^{m(d)}=-2{\rm Im} \Sigma_{b(d)}(0) $ is the maximum optomechanically induced damping rate in which the self-energies $\Sigma_{b}$ and $\Sigma_{d}$ are, respectively, given by Eqs.~(26b) and (26c) of Ref.~\cite{aliDCE3}. Furthermore, $ \mathcal{C}_m $($ \mathcal{C}_d $) is the collective optomechanical cooperativity which is given by 
\begin{eqnarray}\label{Cmd}
	&&  \mathcal{C}_{m(d)}=  \mathcal{C}_{0(1)} \frac{1+ \mathcal{C}_{1(0)} - \xi_{d(m)}^2}{(1+ \mathcal{C}_{1(0)}-\xi_{d(m)}^2)^2- \xi_{d(m)}^2  \mathcal{C}_{1(0)}^2} ~.
\end{eqnarray}

The maximum values of modulation amplitudes are obtained self-consistently from Eq.~(\ref{stability}), as $ \lambda^{max}_{m(d)}=\frac{\gamma_{m(d)}}{2} \left[1+ \mathcal{C}_{m(d)}\right] $. It should be emphasized that so far as the modulation amplitudes are less than their maximum values ($ \xi_{m(d)}^{max}=2 \lambda^{max}_{m(d)}/\gamma_{m(d)}$) the system is stable and the approximations we have made are valid because the system is in the weakly interacting regime.

To find the negativity condition or find the \textit{optimized} paramps $ \xi_m^{\rm opt} $ and $ \xi_d^{\rm opt} $, for example in on-resonance frequency, one should simultaneously solve inequality $4 A_N/A_D=M<0 $ together with the stability condition of Eq.~(\ref{stability}); $ \xi_{m(d)} \le 1+\mathcal{C}_{m(d)} $. In other words, to find the paramps corresponding to the negative CPSF, one should require both inequalities 
\begin{eqnarray} \label{neagitivitycondition}
	&& 4 \frac{A_N(\xi_m,\xi_d)}{A_D(\xi_m,\xi_d)}= M< 0, \quad \&  \qquad \xi_{m(d)} \le 1+\mathcal{C}_{m(d)}.
\end{eqnarray}
Note that $ M $ ($ M > M_{\rm max} $ where $ M_{\rm max} $ is the maximum achievable negativity) is an arbitrary negative value which should satisfy the negativity conditions. Furthermore, the maximum negativity $ M_{\rm max} $ should also satisfy the negativity conditions (\ref{neagitivitycondition}).

Let us consider a special case where just one of the mechanical modes is parametrically driven while the other one remains unmodulated, i.e., $\xi_m\neq 0$ and $ \xi_d=0 $. In this case, the on-resonance CPSF reads as
\begin{eqnarray}
	&& \mathcal{A} (0) \vert_{\xi_d=0}= \frac{4}{\kappa} \frac{1}{1+\mathcal{C}_1} \frac{\xi_m^{max}-\xi_m^2}{(\xi_m^{max})^2-\xi_m^2}, \label{CPSF2}
\end{eqnarray}
where $ \xi_m^{max}=1+\mathcal{C}_0/(1+\mathcal{C}_1) $. As is evident from Eqs.~(\ref{CPSF1}) and (\ref{CPSF2}), the denominators are always positive while the numerators can take negative values. For example, in Eq.~(\ref{CPSF2}) if the modulation amplitude lies in the interval  $ \sqrt{\xi_m^{max}} \le \xi_m \le \xi_m^{max}  $ where the strong-coupling regime is avoided, the system is stable while the CPSF becomes negative near the resonance frequency. 
Note that in the absence of the parametric modulation, the cavity photon spectral function is given by
\begin{eqnarray} \label{A_0}
	&&  \mathcal{A}(0) \vert_{\xi_{m,d}=0}=\frac{4}{\kappa} \frac{1}{1+\mathcal{C}_0} ,
\end{eqnarray}
which never becomes negative.

\begin{figure} 
	\includegraphics[width=6.5cm]{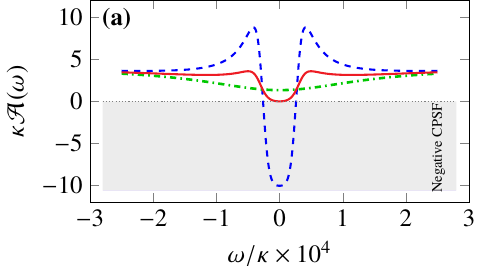}
	\includegraphics[width=6.5cm]{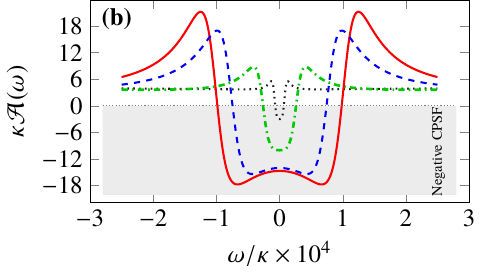}
	\centering
	\caption{Behavior of the CPSF, $ \kappa \mathcal{A}[\omega]  $, vs normalized frequency $ \omega/\kappa $ for the case of standard OMS with a single MO ($ \mathcal{C}_1=\lambda_d=\gamma_d=0 $) which is parametrically driven via the coherent-time modulation of the spring coefficient. (a) For the fixed cooperativity $ \mathcal{C}_0=2 $ and different values of relative amplitude paramp $ \xi_m=0, \sqrt{\xi_m^{\rm max}}, 0.9 \xi_m^{\rm max} $, respectively, corresponding to the green dot-dashed, red solid and blue dashed lines. (b) for the fixed relative amplitude paramp $  \xi_m/\xi_m^{\rm max}=0.9 $ and different values of cooperativity $ \mathcal{C}_0=0.5,2,6,8 $ corresponding, respectively, to the black-dotted, green-dot-dashed, blue-dashed and red-solid lines. Here, we have set $ \kappa/\gamma_m=10^4 $. Note that in the case of a single driven MO we have $ \xi_m^{\rm max}=1+\mathcal{C}_0 $.
	}
	\label{fig2}
\end{figure}

Fig. (\ref{fig2}) shows the effect of modulation on the CPSF in the case of an OMS with a single MO, when $ \mathcal{C}_1=\lambda_d=\gamma_d=0 $. In Fig.~\ref{fig2}(a), we have fixed the cooperativity and plotted the CPSF for different values of the relative amplitude $ \xi_m/\xi_m^{\rm max} $. The dips of all the curves in Fig.\ref{fig2}, which are appeared at the resonance frequency $\omega=0$ corresponding to $\omega_{pc}=\omega_m$, are just the OMIT windows \cite{OMIT1,OMIT2,OMIT3,vitaliOMIT,marquardtOMIT,XiongOMIT,shahidaniOMIT2013,dalafiOMIT2022}. As is evident, in the absence of modulation (green dot-dashed curve), $ \xi_m=0 $,the CPSF remains positive at all frequencies. However, by turning on the parametric drive and adjusting the modulation amplitude in the domain  $ \xi_m \ge \sqrt{\xi_m^{\rm max}} $ the dip of the OMIT window takes negative values at frequencies near the resonance frequency which is the indication of a gain in the system (blue dashed line). This negativity can be controlled by the amplitude of modulation and the amount of the negativity increases when the strength of the paramp increases until both conditions (\ref{neagitivitycondition}) are satisfied in the stable regime. To examine the effect of cooperativity on the behavior of the CPSF, we have plotted in Fig.~\ref{fig2}(b) the CPSF for different values of  cooperativity and a fixed value of the relative amplitude $ \xi_m/\xi_m^{\rm max} $. 
As is seen, the larger the cooperativity $ \mathcal{C}_0$, the more negative the CPSF occurring over a broader range of frequencies.
Moreover, as the cooperativity increases the two symmetric sidebands in the CPSF, arising from the normal mode splitting (NMS), become more prominent. It means that in an OMS with a single mechanical mode the bandwidth of the OMIT window cannot be controlled easily at a fixed value of negativity. The reason is that it can be manipulated just by the system cooperativity $ \mathcal{C}_0$ which not only changes the system negativity but also makes the system regime change from OMIT to NMS. Another feature illustrated in Fig.~\ref{fig2}(b) is that, with increasing $ \mathcal{C}_0$, the maximum negative value of the CPSF occurs at off-resonance frequencies (see blue-dashed and red-solid lines in Fig.~\ref{fig2}(b)).

\begin{figure} 
	\includegraphics[width=6.55cm]{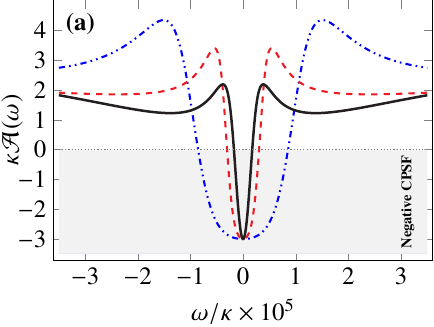}
	\includegraphics[width=6.7cm]{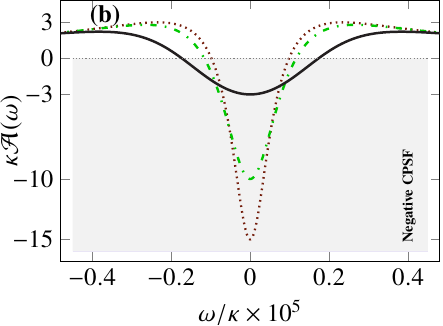}
	\centering
	\caption{Frequency dependence of the CPSF for an OMS with two parametrically driven mechanics with cooperativities $ \mathcal{C}_0=1 $ and $ C_1=0.5 $ and damping rates $ \kappa/\gamma_m=10^4 $ and $ \gamma_m/\gamma_d=1 $. (a) CPSF for three different modulation paramps $(\xi_m,\xi_d)$ corresponding to the fixed negativity value $ M=-3 $. The black-solid, blue double-dot-dashed, and the red loosely-dashed lines correspond, respectively, to the optimized paramps ($ \xi_m= 1.275$, $ \xi_d=0.768 $); ($ \xi_m=1.502 $, $ \xi_d=0 $) and ($ \xi_m=0 $, $ \xi_d=1.199 $). 		
		(b) CPSF for different values of negativity: $ M=-3 $ (the black-solid line) with optimized paramps ($ \xi_m=1.275 $, $ \xi_d=0.768 $); $ M=-10$ (the green dot-dashed line) with optimized paramps ($ \xi_m=1.286 $, $ \xi_d=0.784 $); and $ M=-15 $ (the brown-dotted line) with optimized paramps ($ \xi_m=1.288 $, $ \xi_d=0.787 $. The optimized paramps have been obtained by solving Eqs.~(\ref{neagitivitycondition}) numerically. }	
	\label{fig3}
\end{figure}

In Fig.\ref{fig3}, we have demonstrated the frequency dependence of CPSF for an OMS with two mechanical modes with cooperativities $ \mathcal{C}_0=1 $ and $ C_1=0.5 $ and damping rates $ \kappa/\gamma_m=10^4 $ and $ \gamma_m/\gamma_d=1 $. Fig.\ref{fig3}(a) shows the CPSF behavior for three different ways of modulations corresponding to the fixed negativity value of $ M=-3 $. The black-solid shows the situation where both the mechanical modes are modulated at their optimized values of $ (\xi_m=1.288 $, $ \xi_d=0.787) $ while the blue double-dot-dashed, and the red loosely-dashed lines correspond, respectively, to ($ \xi_m=1.502 $, $ \xi_d=0 $) and ($ \xi_m=0 $, $ \xi_d=1.199 $), where just one of the mechanical modes is modulated. As is seen,  when both mechanical modes are parametrically driven the OMIT window becomes very narrow in comparison to the situation where just one of them is modulated. It means that, just by controlling the modulation powers (paramps), one can manipulate the bandwidth of optomechanical transparency gain and switch from narrow-band to broad-band and vice versa with the same negativity in the OMIT regime. Therefore, one of the most important advantageous aspects of the present OMS with two modulated mechanical modes is that it can be used as a switchable narrow-band and broad-band filter in comparison to an OMS with a single mechanics \cite{optomechanicswithtwophonondriving} whose bandwidth is not controllable. It should be reminded that in the case of an OMS with a single mechanical mode, resonance bandwidth could be changed through the cooperativity. However, changing the cooperativity not only changes the system negativity but also will make the system regime be changed from OMIT to NMS (see the results of Fig.\ref{fig2}). On the other hand, as has been shown in Fig.~\ref{fig3}(b), one can achieve different values of negativity in the stable regime by slowly tuning and increasing both paramps whose optimized values are determined by solving Eqs.~(\ref{neagitivitycondition}) numerically. In short, both the magnitude of the negativity and the OMIT bandwidth can be controlled independently in an OMS with two mechanical modes.

\section{Experimental discussion} \label{sec_experiment}
{\color{black} In order to show how the results demonstrated in Fig.\ref{fig3} can be realized in an experimental setup, it should be noted that for every special experimental setup consisting of an OMS having two mechanical modes with specified optomechanical couplings $(g_0,G_0)$ and mechanical damping rates ($ \gamma_m,\gamma_d $), the values of cooperativities can be controlled by the coupling laser power because the cooperativities depend on the optical mean-field of the cavity. On the other hand, the resonance condition in the red-detuned regime, i.e., $\Delta_0=\omega_m=\omega_d$, leads to the following second order algebraic equation for the coupling laser frequency
	\begin{equation}\label{wcal}
		\omega_c^2-(\omega_0-\omega_m)\omega_c+\frac{2\kappa_{e} P_c(g_0^2+G_0^2)}{\hbar\omega_m(\kappa^2/4+\omega_m^2)}=0,
	\end{equation}
	which gives us the true coupling laser frequency versus the coupling laser power. It should be noted that in order to have real roots the following condition should be satisfied by the coupling laser power
	\begin{equation}\label{Pcc}
		P_c\leq  P_c^{\rm max}= \hbar\omega_m\frac{\kappa^2/4+\omega_m^2}{8\kappa_{e}(g_0^2+G_0^2)} (\omega_0-\omega_m)^2.
	\end{equation}
	Therefore, for every specified value of the coupling laser power satisfying Eq.(\ref{Pcc}) there will be two real roots for $\omega_c$ which just one of them, that satisfies the system stability conditions (the Routh-Hurwitz criteria), is acceptable. Once the coupling laser power and its true frequency are determined, the system cooperativities are also determined. 
	Then, for any specified values of the system cooperativities and a required negative value of spectral function ($M<0$), one can obtain the right values of modulation amplitudes using the system stability conditions of Eq.(\ref{stability}).
	Since there is an upper limit for the coupling laser power $P_c$ based on Eq.(\ref{Pcc}), there is an upper limit for each cooperativity so that $0<\mathcal{C}_0<\frac{4 g_0^2}{\kappa\gamma_m}\bar a^2_{max}$, $0<\mathcal{C}_1<\frac{4 G_0^2}{\kappa\gamma_m}\bar a^2_{max}$, where $\bar a_{max}$ is the maximum value of the optical mean-field corresponding to the maximum of $P_c$, i.e.,  $ \bar a^2_{max}=  \dfrac{\kappa_{e} P_c^{\rm max}}{\hbar \omega_c (\omega_m^2 +\kappa^2/4)} $. For every allowed value of the cooperativities there exist optimized values of modulation amplitudes that are determined by Eq.(\ref{stability}) and give us a specified negative value of $M$. Besides, the modulation amplitude are not allowed to be greater than their maximum values $(\xi_m^{max},\xi_d^{max})$ based on the stability condition. }

\begin{figure} 
	\includegraphics[width=6.5cm]{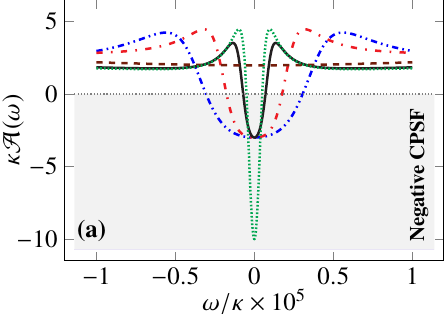}
	\includegraphics[width=6.5cm]{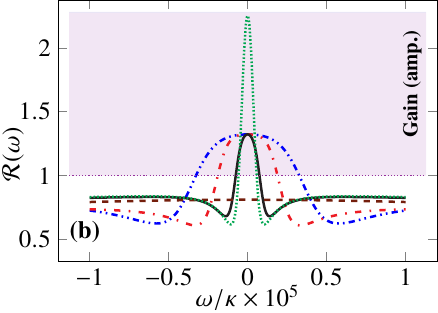}
	\centering
	\caption{ (a) The CPSF and (b) power reflection coefficient versus normalized frequency based on the experimental data given in Ref.\cite{piergentili2021two}. If the coupling laser power has been fixed at  $ P_c \simeq 0.067 \mu  $W the optomechanical cooperativities takes the feasible values $ C_0 \simeq 0.69 $ and $ C_1 \simeq 0.33$. The brown dashed-line shows the off-modulation case ($ \xi_m=\xi_d=0 $). The black-solid, red dot-dashed, and blue double-dot dashed lines representing $ M=-3 $, have optimized paramps given by ($ \xi_m \simeq 1.390 $, $ \xi_d \simeq 0.879 $), ($ \xi_m \simeq 1.382 $, $ \xi_d=0 $) and ($ \xi_m =0  $, $ \xi_d \simeq 1.151 $), respectively. The green-dotted line representing $ M=-10 $ have optimized paramps $ \xi_m \simeq 1.253 $, $ \xi_d \simeq 0.737 $. Here in (b) we have assumed that $ \kappa^{\prime}= 0.1 $.
	}
	\label{fig4}
\end{figure}

In the following, we will discuss our results based on the above explanations and the experimentally feasible data presented in Ref.\cite{piergentili2021two}. For this purpose, we consider an optical cavity with length $ L=53.571(9) \mu $m consisting of two $\rm Si_{3}N_{4}$ membranes in the middle with equal natural frequencies $ \omega_m\simeq\omega_d \simeq 2\pi \times 230 $kHz, and different damping rates $ \gamma_m\simeq 2\pi \times 1.64 $Hz and $ \gamma_d \simeq 2\pi \times 9.37  $Hz. The cavity, which has a finesse of $  \mathcal{F} = 12463(13) $ corresponding to an optical decay rate $ 2\kappa \simeq 2\pi \times 134 $ kHz, is driven by a coupling laser with a central wavelength of $ \lambda_c=1064 $nm. The single-photon optomechanical couplings are $ g_0 \simeq 2 \pi \times 0.4225 $Hz and $ G_0 \simeq 2 \pi \times 0.6965  $Hz. The external cavity loss and intrinsic (internal) cavity decay rate are, respectively, $ \kappa_{e} \simeq 2 \pi \times 50.35$kHz and $ \kappa_i \simeq 2 \pi \times 8.35 $kHz.

In Fig.\ref{fig4}, we have plotted the CPSF as well as the power reflection coefficient $\mathcal{R}$ versus the normalized frequency. Here, it has been assumed that the coupling laser power is fixed at $ P_c \simeq 0.067 \mu  $W which leads to a mean cavity photon number $ \bar n_{\rm cav}=\bar{a}^2 \simeq 105677 $. Then, using Eq.~(\ref{wcal}) and the stability condition the right value of $\omega_c$ is determined and consequently the experimentally feasible cooperativities are obtained as $ C_0 \simeq 0.69 $ and $ C_1 \simeq 0.33$.

As is seen from Fig.\ref{fig4}, just by manipulating the modulation amplitudes, one can control both the magnitude of negativity (the system gain) and the resonance bandwidth of the OMIT window independently. For example the  black-solid, red dot-dashed, and blue double-dot dashed lines show the situation in which the MOs are so modulated that the negativity has been fixed at $M=-3$. For this value of negativity the power reflection coefficient gets larger than unity which means the manifestation of a gain in the system. Interestingly, by changing the modulation amplitudes and fixing them in another optimized point the value of negativity can reach $M=-10$ (the green-dotted line of Fig.\ref{fig4}(a)) which corresponds to a power reflection coefficient larger than 2 (the green-dotted line in Fig.\ref{fig4}(b)). It means that under these conditions the input signal can be amplified more than $200\%$ for an OMS with $\kappa'=0.1$ . The other important point is that by choosing different configurations of the modulation amplitudes (for example by turning on and off one of them or by turning on both of them and fixing them at an optimized point) one can easily change the bandwidth of the OMIT window without perturbing the negativity value (the gain of the system) and without changing the system regime. In this way, the present two-mode OMS can be used as a quantum linear amplifier and also as a controllable narrow-band and broad-band filter.

Therefore, the importance of the spectral function negativity is that it leads to a special kind of OMIT, the so-called \textit{extraordinary} OMIT where the intensity of the probe field in the output becomes greater than its input intensity at the center of the OMIT window. Such a modulated OMS can be applied in future quantum technologies as an \textit{optical transistor} that can amplify the input optical signal.

Although adding an extra mechanical mode to the system causes the experimental setup construction to be more difficult and costly and makes more noises enter the system \cite{aliDCEforcesenning}, it plays a very important role in quantum amplification and control \cite{aliDCEsqueezing} as well as optomechanically  photon/phonon generation mechanism via the dynamical Casimir effect (DCE) \cite{aliDCE3}. In fact, based on the theory of linear quantum amplifiers \cite{quantumnoise}, in order to improve the functionality of a linear amplifier it is necessary to add more degrees of freedom to the system so that the input signal is amplified more effectively. However, the price to pay for introducing extra degrees of freedom will be the manifestation of some added noises to the input signal.  It should be noted that vanishing of the quantum noises in the mean value of output cavity field (as is seen in Eq.(\ref{eout})) does not mean a noiseless amplification in the present system because the thermal noises as well as the imprecision shot noise and radiation pressure backaction noise are always appeared in the power spectrum of the cavity output and therefore there is always a finite value of signal to noise ratio. Nevertheless, there exist methods which can reduce the added noises of measurement. As has been shown in Ref.\cite{aliDCEforcesenning} using an OMS with two modulated mechanical modes (where the second mode has been considered as the collective mode of a BEC), the signal to noise ratio can be enhanced very considerably in comparison to an OMS with just one modulated mechanical mode. In an ultraprecision quantum measurement like that studied in Ref.\cite{aliDCEforcesenning}, the role of added noises becomes very important because the weak input signal is not detectable unless the added noises are suppressed considerably. 

In the following, we present the physical interpretation of the negative CPSF in our system by comparing to the OPA or DPA. 

\section{Comparison to the OPA  and Physical interpretation \label{sec.comparisontotheOPA}}

We are interested in making a comparison between the CPSF of the present system and that of an OPA since our system is effectively analogous to an OPA with a time-dependent amplitude which is responsible for the frequency-dependent squeezing coefficient in the Fourier space \cite{aliDCE3,aliDCEsqueezing,aliDCEforcesenning,optomechanicswithtwophonondriving}. For a \textit{detuned} OPA, where the pump field frequency ($ \omega_p/2 $) is not equal to the cavity frequency ($\omega_c$), the Hamiltonian is given by
\begin{eqnarray} \label{H_detunedOPA}
	&& \hat H^{(\rm OPA)}_{(\rm det)} = \hbar\omega_c \hat a^\dag \hat a + \frac{i}{2} (\lambda \hat a^{\dag 2} e^{-i\omega_p t} - \lambda^\ast \hat a^2 e^{+i\omega_p t}) ,
\end{eqnarray}
where in a frame rotating at $ \omega_p/2 $ the Hamiltonian takes the following form
\begin{eqnarray} \label{H_detunedOPA2}
	&& \hat H^{(\rm OPA)}_{(\rm det)} = -\hbar\Delta_p \hat a^\dag \hat a + \frac{i\hbar}{2} (\lambda \hat a^{\dag 2} - \lambda^\ast \hat a^2) ,
\end{eqnarray}
with $ \Delta_p=\omega_p/2-\omega_c $.

By solving the equations of motion in Fourier space we find that our system resembles an OPA with an effective frequency-dependent parametric drive $ \lambda \to \tilde \lambda_a(\omega) $ and with a non-zero cavity self-energy $ \Sigma_a(\omega) $ [see Eqs.~(\ref{a_omega_eq}) and (\ref{lambda_all})]. The susceptibility matrix for the detuned OPA is given by
\begin{eqnarray} \label{chiOPAdetuned}
	&& \boldsymbol{\chi}^{(\rm OPA)}_{(\rm det)}(\omega)= \frac{1}{(\kappa/2-i \omega_+) (\kappa/2-i \omega_-)-\vert \lambda \vert ^2} \left( \begin{matrix}
		{\kappa/2-i\omega_-} & {\lambda} \\
		{\lambda^\ast} & {\kappa/2-i \omega_+} \nonumber  \\
	\end{matrix} \right), \nonumber  \\
\end{eqnarray} 
with $ \omega_{\pm}= \omega \pm \Delta_p $, which leads to the following retarded Green's function
\begin{eqnarray} \label{greenOPA}
	&& G^{(\rm OPA)}_{aa^\dag}(\omega)=-i \chi^{(\rm OPA)}_{aa}(\omega)  = -i \frac{[(\kappa/2 - i (\omega-\Delta_p)] [\kappa^2/4 -\vert \lambda \vert^2 - (\omega^2-\Delta_p^2)+i \omega \kappa ]}{[\kappa^2/4 -\vert \lambda \vert^2 - (\omega^2-\Delta_p^2)]^2 +\omega^2 \kappa^2} . 
\end{eqnarray}
Therefore, the CPSF for the detuned OPA can be found as follows
\begin{eqnarray} \label{CPSF-OPA-detuned}
	&& \mathcal{A}^{(\rm OPA)}_{(\rm det)}(\omega)= - 2{\rm Im } G^{(\rm OPA)}_{aa^\dag}(\omega) =+\kappa \frac{[\kappa^2/4 -\vert \lambda \vert^2 - (\omega^2-\Delta_p^2) ] + 2\omega (\omega -\Delta_p) }{[\kappa^2/4 -\vert \lambda \vert^2 - (\omega^2-\Delta_p^2)]^2 + \omega^2 \kappa^2}. 
\end{eqnarray} 
The stability condition, $ \rm det \boldsymbol{\chi}^{(\rm OPA)}_0 >0$, requires that the poles of the susceptibility lie in the lower half-plane. For the detuned-OPA the stability condition becomes $ \vert \lambda \vert^2 < \kappa^2 /4+ \Delta_p^2 $. In order to examine the negativity of the CPSF for an OPA, let us rewrite Eq.~(\ref{CPSF-OPA-detuned}) as follows
\begin{equation} 
	\mathcal{A}^{(\rm OPA)}_{\rm (det)}(\omega)=\kappa \frac{F(\omega)}{[\kappa^2/4 -\vert \lambda \vert^2 - (\omega^2-\Delta_p^2)]^2 + \omega^2 \kappa^2},  \label{Arewite}
\end{equation}
with $ F(\omega)= \omega^2 -2\Delta_p \omega +S$ and $ S= \kappa^2 /4+ \Delta_p^2 - \vert \lambda \vert^2  >0 $ according to the the stability condition. 
It follows from Eq.~(\ref{Arewite}) that the stable negativity in CPSF occurs if $ F(\omega)<0 $ for any positive frequency $ \omega>0 $. Evidently, for the on-resonance condition we have $ F(0)=S $ which is always positive in the stability region and thus the spectral function never becomes negative in on-resonance frequency ($ \mathcal{A}^{(\rm OPA)}_{\rm (det)}(\omega=0) >0 $).

Let us now see what happens with the off-resonance frequencies for the OPA or detuned OPA. 
In this case, to find the negativity condition in off-resonance frequencies for the CPSF, one must solve inequality $ F(\omega>0) < 0 $ which yields the constraint $ \kappa^2 /4 < \vert \lambda \vert^2 < \kappa^2 /4+ \Delta_p^2 $. It is clear that the CPSF for nondetuned OPA, i.e., when $ \Delta_p=0 $, never becomes negative in off-resonance frequencies. 
In other words, as the detuned OPA is stable the CPSF can be negative in off-resonance frequencies  $ \Delta_p -\Delta \omega \le \omega \le \Delta_p +\Delta \omega $ where $ \Delta \omega=\sqrt{\Delta_p^2-S}= \sqrt{\vert \lambda \vert^2 - \kappa^2/4} $ when $ \kappa^2 /4 < \vert \lambda \vert^2 < \kappa^2 /4+ \Delta_p^2 $.

\begin{figure} 
	\includegraphics[width=9cm]{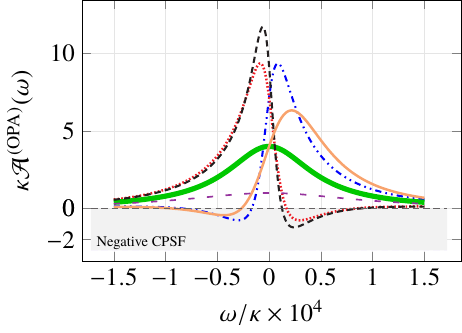}
	\centering
	\caption{ CPSF for the OPA vs normalized frequency $ \omega/\kappa $ for different values of $ \xi_k= \frac{\lambda}{\kappa/2} $ and $ \Delta_k= \Delta_p/\kappa $. The green thick-solid and purple loosely-dashed lines refer to zero detuning $ \Delta_p=0 $, respectively for $ \xi_k=0 $ and $ \xi_k=1$. Red densely-dotted and blue double-dot dashed lines refer to the same paramp $ \xi_k=1.2 $, respectively, for $ \Delta_k=0.5 $ and $ \Delta_k=-0.5 $. Black densely-dashed and orange-solid lines refer to ($ \xi_k=1.25 $, $ \Delta_k=0.5 $) and ($ \xi_k=1.2 $, $ \Delta_k=-0.6 $), respectively.
	}
	\label{fig5}
\end{figure}

In Fig.~(\ref{fig5}), we have plotted the CPSF for the nondetuned and detuned OPA (Eq.(\ref{CPSF-OPA-detuned})). As is evident for the case of nondetuned OPA (see the green thick-solid and purple loosely-dashed lines), the CPSF never becomes negative and has a positive peak at the resonance frequency which decreases with increasing the effective paramp ($ \xi_k=2\lambda/\kappa $) to its maximum value $ \xi_k=1 $. Note that the case of zero detuning and absence of paramp corresponds to an empty cavity subjected to dissipation. However, for the detuned OPA, the CPSF can be negative in off-resonance frequencies, as expected. 

For a fixed value of the effective paramp $ \xi_k $ but opposite values of effective detuning $ \Delta_k $ (see red densely-dotted and blue double-dot dashed lines in Fig.~(\ref{fig5})) the height of the peaks in CPSF is the same, while their locations are dependent on the sign of the detuning $ \Delta_k $. Moreover, the spectral function exhibits negativity over a range of positive (negative) frequencies for positive (negative) detuning $ \Delta_k $. On the other hand, for a fixed value of the effective detuning $ \Delta_k $ the spectral function negativity is enhanced with increasing the paramp $ \xi_k $ (see red densely-dotted and black densely-dashed lines in Fig.~(\ref{fig5})). It is also seen from Fig.~(\ref{fig5}) (see orange-solid and blue double-dot dashed lines) that for a fixed value of the effective paramp $ \xi_k $ a larger value of detuning $ \Delta_k $ leads to smaller negativity of the spectral function.

Here, it is worth remarking that unlike the detuned OPA, the parametrically driven OMS considered in this paper can exhibit stable negative spectral function at on-resonance frequency ($ \omega=0 $). This is because by controlling the system parameters, in particular the effective induced paramp, the spectral function negativity can be achieved even at $ \omega=0 $ due to the negative ECDR (see Eq.(\ref{A2})), which will be explained in the following.

Let us now illustrate more clearly the physics behind the spectral function negativity in the stable regime in the OPA and in our parametrically-driven OMS, and thus, compare them together. As was shown, the stable negativity in the cavity spectral function of an OPA occurs only in the detuned regime ($ \Delta_p \ne 0 $) due to the extended stability region. However, based on Eq.(\ref{A2}) the negativity of the CPSF in our system originates from the negativity of the frequency dependent ECDR.
For comparison, let us first determine the ECDR for a detuned OPA whose self-energy matrix, $ \boldsymbol{\Sigma}_{(\rm det)}^{(\rm OPA)}(\omega)= i (\boldsymbol{\chi}_0^{-1} - (\boldsymbol{\chi}_{(\rm det)}^{{\rm (OPA)}})^{-1}) $, by using Eq.~(\ref{chiOPAdetuned}), is obtained as
\begin{eqnarray} \label{detuned-selfenergy}
	&& \boldsymbol{\Sigma}_{\rm (det)}^{\rm (OPA)} (\omega)= i \left( \begin{matrix}
		{0} & {\lambda} \\
		{\lambda^\ast} & {0}\\
	\end{matrix} \right),
\end{eqnarray}
which is clearly frequency-independent and purely off-diagonal. Therefore, $ (\kappa_{\rm eff})^{(\rm OPA)}_{(\rm det)} = \kappa >0$ indicating that the EDCR for a detuned OPA is always positive.

Despite the OPA, we will show that in our system the ECDR can be negative, $ \kappa_{\rm eff} (\omega) < 0 $, which reveals the physics behind this negativity in comparison to detuned OPA.
In fact, the negative $ \kappa_{\rm eff}(\omega) $ in our system is because of its frequency dependency through frequency-dependency of the cavity self-energy $ {\rm Im} \Sigma_a(\omega) $ due to $ \tilde \lambda_a(\omega) $ (induced-frequency-dependent paramp) which originates from the coherent time-modulations of the mechanical spring coefficients of the MOs.

Now, we return to our system and calculate its ECDR. For this purpose, let us first make a comparison between the paramps for an ordinary OPA and the optomechanical system under consideration. In an ordinary OPA when its parametric drive strength satisfies the relation $ (\lambda)^\ast =\lambda^\ast$, i.e., the real paramp, the interaction is called a \textit{coherent interaction} \cite{optomechanicswithtwophonondriving,aliDCE3}.   
Generally in our parametrically-driven system, a frequency-dependent effective parametric drive like $ \tilde \lambda_a(\omega) $ satisfies the relation $(\tilde \lambda_a(\omega))^\ast=\tilde \lambda_a^\ast(-\omega) $. 
If $ \tilde \lambda_a(\omega) $ satisfies the relation $(\tilde \lambda_a(\omega))^\ast=\tilde \lambda_a(\omega) $ the system behaves like an ordinary OPA in a coherent regime. In this situation, $ \tilde \lambda_a(\omega) $ can be considered as an effective coherent interaction strength whose real and imaginary parts might lead to the coherent and dissipative behaviors. In this way, two different regimes can be distinguished depending on which part is larger than the other. In the so-called \textit{coherent} regime where \cite{aliDCE3}
\begin{eqnarray}  \label{coherentregime}
	&&  \lvert \frac{{\rm Im} \tilde \lambda_a(\omega)}{{\rm Re} \tilde \lambda_a(\omega)} \vert \ll 1 
\end{eqnarray} 
is satisfied (for example, it can be satisfied in the largely different cooperativities regime with nonzero modulation such that $ \xi_{m(d)} > 1$), the coherent term is dominant otherwise the dissipative term is dominant.

Because of the frequency dependence of $ \tilde \lambda_a(\omega) $ and $ \Sigma_a(\omega) $, we cannot directly map our system to an OPA. However, in the on-resonance case, the effective susceptibility matrix becomes exactly similar to an OPA but with the effective parametric strength $ \tilde \lambda_a(0) $ instead of $ \lambda $ and the effective cavity damping rate $ \kappa_{\rm eff}(0)= \kappa-2{\rm Im} \tilde \Sigma_a(0) $ (with $\kappa_{\rm opt}=-2{\rm Im} \tilde \Sigma_a(0) $) instead of $ \kappa$.

In the stable regime and when the coherent-regime condition (\ref{coherentregime}) is satisfied the optomechanically-induced cavity damping rate, $ \kappa_{\rm opt}=-2{\rm Im} \tilde \Sigma_a(0) <0 $ becomes negative. Consequently, the effective cavity damping rate $ \kappa_{\rm eff}= \kappa + \kappa_{\rm opt}(\omega)$ decreases, and can even become negative by controlling the system parameters. As a numerical example, for the parameters given in Fig.~(\ref{fig3}) corresponding to $ M=-3 $, we obtain $ \kappa_{\rm eff}/\kappa\simeq -1.3 $. Thus, within the analogy between our system and OPA in the on-resonance frequency, the negative CPSF ($ \mathcal{A}<0 $) is responsible for the negative ECDR ($  \kappa_{\rm eff} <0 $).
Therefore, the on-resonance CPSF ($ \mathcal{A}(0) $) for our system can be found by simply making replacements $ \kappa \to \kappa_{\rm eff}(0) $ and $ \lambda \to \tilde \lambda_a(0) $ in $ A^{(\rm OPA)}_{\rm (det)}(0) \vert_{\Delta_p=0} $. 
Note that this analogy cannot be seen as a coincidence, it only shows that the CPSF in our system looks exactly like that of a non-equilibrium OPA which can provide the link between the negative CPSF and associated negative photon temperature in our system.

\section{Concluding remarks, discussion and outlooks \label{summary}}

In summary, we have theoretically proposed and investigated an experimentally feasible scheme to generate \textit{negative} CPSF in an OMS with two parametrically modulated MOs in the red-detuned and weak coupling regimes. Using the GLRT we have calculated CPSF and have shown that under special conditions such a system exhibits an optomechanical induced gain (OMIG) which leads to the amplification of input probe with a controllable bandwidth. The advantage of an OMS with two parametrically driven MOs is that it provides more controllability to achieve negative CPSF in the stable regime, and also enables us to manipulate the OMIT bandwidth in comparison to a system with a single mechanical mode. Moreover, this negativity never occurs in the standard OMS.

The comparison of our parametrically-driven OMS with the conventional OPA and detuned OPA shows that the CPSF negativity never occurs in the conventional OPA in the stable regime while it occurs in the detuned OPA in the stable regime. However, the nature of the negativity in our system is totally different. This negativity originates from the negative ECDR because of its frequency dependence while the ECDR in the detuned OPA is zero and the CPSF negativity stems from the extension of the stability condition of the susceptibility matrix due to the presence of the cavity detuning.

We would like to remind that if the two mechanical modes are not degenerate then the resonance condition of the red-detuned regime, i.e., $ \Delta_0 = \omega_m = \omega_d $, which is the crucial condition of the present scheme for the occurrence of OMIT, is no longer satisfied. In other words, since for non-degenerate mechanical modes the resonance condition in the red detuned regime cannot be satisfied, the most interesting phenomena like OMIT and optomechanical gain do not occur. Nevertheless, other phenomena such as Fano resonance may be observable. However, in this case the Hamiltonian becomes time-dependent so that a straightforward analytical solution is no longer possible and one needs to combine Floquet and Lyapunov techniques \cite{radimFloquet2020} to transform the original time-dependent problem into a time-independent one.

As an outlook, in the system under consideration the real part of the cavity retarded Green's function which is responsible for the intracavity field \textit{dispersion}, effective \textit{frequency-dependent} intracavity \textit{refractive} index could be engineered to achieve a switchable\textit{ slow- and fast-light} by controlling the paramps at fixed optomechanical parameters and gain (the same negativity and power reflection).

Furthermore, negative CPSF at the positive frequencies, i.e., $ \mathcal{A}(\omega>0) <0 $, is responsible for a stationary population inversion between stationary eigenstates separated by $ \hbar \omega $ for a time-independent Hamiltonian in a time-independent state \cite{optomechanicswithtwophonondriving,quantumnoise,greenScarlatella1}. Although our system as an open quantum system with time-dependent Hamiltonian interacts with environment, but the negative CPSF is still responsible for the population inversion. Therefore, it is worthwhile to address the effective temperature of the intracavity photons which can be obtained via the quantum noise properties of the system. As discussed in Refs.~\cite{optomechanicswithtwophonondriving,quantumnoise,greenScarlatella1}, the effective temperature can be defined using the quantum noise properties of photons field by comparing the size of the classical symmetrized photon correlation function to the size of the spectral function using the so-called Keldysh Green's function \cite{Non-Equilibrium-QFT,greenScarlatella1}. If the system is not in equilibrium, effective temperature is frequency-dependent. In thermal equilibrium, effective temperature is exactly the temperature of the system at all frequencies, while, out of equilibrium, for positive frequencies a negative CPSF necessarily leads to a frequency-dependent NET for the intracavity photons \cite{optomechanicswithtwophonondriving} which can be directly probed by coupling the cavity field weakly to an auxiliary probe qubit with a splitting frequency  equal to the cavity frequency \cite{greenScarlatella1,quantumnoise,optomechanicswithtwophonondriving}. Then, the intracavity photons act as an effective engineered bath for the qubit. Since, the steady-state population inversion of the qubit depends on the effective cavity temperature at cavity frequency \cite{quantumnoise}, thus, a NET for the intracavity photons can be directly translated into a simple population inversion of the qubit. In light of this issue, a particular focus in our future work will be the investigation of realizing a controllable NET for the intracavity photons in the stable regime of a hybrid qubit-OMS with two parametrically driven mechanical modes.

\section*{Acknowledgements}
The authors would like to express their gratitude to the anonymous referees whose valuable comments and fruitful suggestions have improved the paper substantially. AMF would like to thank the Office of the Vice President for Research of the University of Isfahan and ICQT for their supports. AD gratefully acknowledges supports from the Iran National Science Foundation (INSF) under Grant No. 99020597.

\section*{Disclosures}
The authors declare no conflicts of interest.

\section*{Data availability}
All data that support the findings of this study are included within the article.

\appendix
\section{Derivation of the Hamiltonian of MO parametric modulation}\label{apA}
If the MOs are parametrically driven by modulating their spring coefficients at twice of their natural frequencies as $k_{m(d)}(t)= k_{m(d)}+\delta k_{m(d)} \sin(2\omega_{m(d)} t +\varphi_{m(d)}) $ which is equivalent to the modulation of the mechanical frequencies [ see Fig. (\ref{fig1})], then the Hamiltonian of MOs can be written as \cite{optomechanicswithtwophonondriving}
\begin{eqnarray} \label{H_m(t)}
	\hat H_{\rm MO}(t)&=&\sum_{j=m,d}\frac{ \hat p_j^2}{2m_j} + \frac{1}{2} k_j(t) \hat x_j^2,\nonumber\\
	&& =\hbar\omega_m\hat b^{\dagger}\hat b + \hbar\omega_d\hat d^{\dagger}\hat d + \hat H_{mod},
\end{eqnarray}
where 
\begin{eqnarray} \label{H_mod(t)}
	&& \hat H_{\rm mod}(t)= \sum_{j=m,d}\frac{1}{2} \delta k_j \sin(2\omega_j t+\varphi_j) \hat x_j^2,
\end{eqnarray}
is the modulation part of the MOs, and the position operators of the MOs are, respectively, $\hat x_{m}=x_{zp(m)}(\hat b + b^{\dagger})$, and $\hat x_{d}=x_{zp(d)}(\hat d + d^{\dagger})$, 
It should be reminded that we have assumed the two MOs have equal frequencies, and the system is in the red detuned regime of $\Delta_0=\omega_m=\omega_d$. Now, using the rotating wave approximation (RWA) the fast rotating term over time scales longer than $ \omega_m^{-1} $ can be ignored so that the Hamiltonian of Eq. (\ref{H_mod(t)}) is simplified as 
\begin{eqnarray} \label{H_mod}
	&& \hat H_{\rm  mod}(t)= \frac{i \hbar}{2} (\lambda_m \hat b^{\dag 2}  e^{-2i\omega_m t}- \lambda_m^\ast \hat b^2 e^{2i\omega_m t})+ \frac{i \hbar}{2} (\lambda_d \hat d^{\dag 2}  e^{-2i\omega_m t}- \lambda_d^\ast \hat d^2 e^{2i\omega_m t}),
\end{eqnarray}
where $ \lambda_m(d)= \vert \lambda_m(d) \vert e^{-i\varphi_{m(d)}} $ with $ \vert \lambda_m(d) \vert= \delta k_{m(d)} x_{zp(md)}^2 / 2\hbar $. It is because of the fact that in the interaction picture with $H_0=\hbar\omega_m (\hat a^\dag \hat a + \hat b^\dag \hat b +  \hat d^\dag \hat d)$, the above terms becomes time-independent while the others (the deleted ones) become time-dependent which can be ignored in the RWA.

In the context of mechanical frequency modulation there are several theoretical proposals \cite{theoreticalMechanicalModulation1,theoreticalMechanicalModulation2,theoreticalMechanicalModulation3,theoreticalMechanicalModulation4} as well as experimental References \cite{pontin2016dynamical,jaskulaDCEBEC} in which the coherent modulation of the mechanical spring coefficient of the MO or the time modulation of the s-wave scattering frequency of atom-atom interaction of the BEC have been experimentally realized. In addition, the combination of graphene and superconducting cavity is a very promising tool for achieving remarkable adjustment of the optomechanical properties, where the modulation of the mechanical mode has been also achieved by tuning the voltage of the gate electrode for the graphene resonator \cite{graphene1,graphene2,graphene3}.

Recently, some other experimental platforms have been also proposed to investigate the effects induced by the parametrical modulation of mechanical frequency. One of such platforms is based on the cantilever optomechanical system \cite{cantilever2018ModulationExperiment1}, where a harmonically oscillating optical trap (optical force) is applied for modulating the effective frequency of cantilever. 
Another platform is based on levitated optomechanics \cite{nanoparticleModulation2016Experiment,levitated2015ModulationExperiment} where a nanoparticle is confined within a hybrid electro-optical trap formed by a Paul trap within a single-mode optical cavity. In this platform the periodic modulations are generic and occur in optically trapped setups in which the equilibrium point of the oscillator is varied cyclically. Furthermore, the realization of the frequency modulation of the cavity field mode has also been experimentally demonstrated by replacing the smaller Josephson junction of the qubit with a tunable superconducting quantum interference device (SQUID) \cite{cavityModulation1Exp2009,cavityModulation2Exp2017}. 
In such a system the modulation of the cavity field mode has been realize by tuning the magnetic flux in superconducting systems.

\bibliography{negativityreferences}

\end{document}